\documentclass[prb,twocolumn,showpacs,amsmath,amssymb]{revtex4}

\usepackage{graphicx}
\usepackage{dcolumn}
\usepackage{bm}
\usepackage{epsfig}

\begin{document}

\newcommand{\ep}{\varepsilon}
\newcommand{\up}{\uparrow}
\newcommand{\dn}{\downarrow}
\newcommand{\vectg}[1]{\mbox{\boldmath ${#1}$}}
\newcommand{\vect}[1]{{\bf #1}}

\title{Modulating Unpolarized Current in Quantum Spintronics: Visibility of
Spin-Interference Effects in Multichannel  Aharonov-Casher Mesoscopic Rings}

\author{Satofumi Souma and Branislav K. Nikoli\' c}
\affiliation{Department of Physics and Astronomy, University
of Delaware, Newark, DE 19716-2570}

\begin{abstract}
The conventional unpolarized current injected into a {\em quantum-coherent} semiconductor
ring attached to two external  leads can be modulated from perfect conductor to perfect
insulator limit via Rashba spin-orbit (SO) coupling. This requires that ballistic propagation
of electrons, whose spin precession is induced by the Aharonov-Casher phase, takes place
through a single conducting channel ensuring that electronic quantum state remains a pure separable one in the course of transport. We study the fate of such spin interference effects
as more than one orbital conducting channel becomes available for quantum transport. Although
the conductance of multichannel rings, in general, does not go all the way to  zero at any
value of the SO coupling, some degree of current modulation survives. We analyze possible scenarios that can lead to reduced visibility of spin interference effects that
are responsible for the zero conductance at particular values of the Rashba
interaction: (i) the transmitted spin states remain fully coherent, but conditions for
destructive interference are different in different channels; (ii) the transmitted spins
end up in partially coherent quantum state arising from entanglement to the environment
composed of orbital degrees of freedom of the same particle to which the spin is attached.
\end{abstract}

\maketitle

\section{Introduction} \label{sec:introduction}
Recent attempts in spintronics~\cite{spintronics} to harness
electron spin for classical and quantum information processing have
encountered two major challenges: efficient room temperature spin
injection~\cite{injection} into a semiconductor and quantum-coherent
control of spin states.~\cite{kikkawa} For example, a paradigmatic semiconductor
spintronic device, the Datta-Das spin-field-effect transistor~\cite{datta90}
where current passing through a two-dimensional electron gas (2DEG) in
semiconductor heterostructure is modulated by changing the strength of
Rashba SO interaction via gate electrode,~\cite{nitta} requires
both problems to be surmounted. The usage of the Rashba SO coupling
(which arises due to inversion asymmetry of the confining electric potential
for 2DEG) to control spin via electrical means has become one of the most
influential concepts in semiconductor spintronics.~\cite{injection} The injected
current can be modulated in this scheme only if it is fully polarized, while precessing
spin has to remain quantum coherent during propagation between the two ferromagnetic
electrodes. Although spin injection into bulk semiconductors has been demonstrated at
low temperatures,~\cite{ohno} creating and detecting spin-polarized currents in
high-mobility 2DEG has turned out to be a much more demanding task. In addition, the
2DEG region of spin-FET would be very sensitive to the stray fields induced by the
ferromagnetic electrodes.~\cite{johnson}

For devices pushed into the mesoscopic realm,~\cite{meso,carlo_rmt}
it becomes possible to modulate even unpolarized currents by exploiting
spin-dependent quantum interference effects in phase-coherent transport. This
is due to the fact that at low temperature $T \ll 1K$ and at nanoscales full
electron quantum state $|\Psi \rangle \in {\mathcal H}_o \otimes {\mathcal H}_s$
remains pure in the tensor product of orbital and spin Hilbert spaces, respectively.
The conductors in the shape of multiply-connected geometries have been an essential
playground since the dawn of mesoscopic physics~\cite{meso,webb} to explore how
quantum interference effects, involving topological quantum phases~\cite{geo_phase}
acquired by a particle moving in the presence of electromagnetic potentials, leave
signatures on the measurable transport properties. The typical example is a metallic
ring in the magnetic field where transported electron encircling magnetic flux
acquires Aharonov-Bohm topological phase which induced conductance oscillations.~\cite{webb}
The electromagnetic duality entails an analogous effect---a neutral magnetic moment going
around a charged line acquires Aharonov-Casher (AC) phase that can be manifested in a
multitude of ways in mesoscopic transport quantities.~\cite{entin,marthur,aronov,su}

In recently proposed spintronic ring device~\cite{ring} containing the Rashba SO coupling,
the difference between AC phases of opposite spin states traveling clockwise and
counterclockwise around the ring generates spin interference effects that can be observed in
its transport properties. Since the ring conductance directly depends 
on this difference,~\cite{ring,diego} such mesoscopic quantum interference effects can be
exploited to modulate conductance of unpolarized charge transport through a
one-dimensional (1D) ring (attached to two 1D leads) between 0 and $2e^2/h$ by changing
the Rashba electric field via gate electrode covering the structure.~\cite{nitta} 
The attractiveness of such {\em all-electrical} and {\em all-semiconductor}  device comes from 
the fact that evades usage of any ferromagnetic elements or magnetic fields. 
However, its envisaged operation necessitates that quantum transport takes place 
through {\em only} a single Landauer conducting channel.~\cite{carlo_rmt} 
The theoretical analysis thus far has been confined to strictly 1D rings,~\cite{ring,molnar} 
or 2D rings where only the lowest transverse propagating mode is open for quantum transport.~\cite{diego} 
In both of these cases of single-channel transport similar pattern of complete conductance modulation 
between 0 and $2e^2/h$ is found, even though higher unoccupied modes in 2D rings can affect 
the lowest open channel.~\cite{decoherence,hausler}

On the other hand, despite advances in nanofabrication technology, it is quite
challenging to fabricate single channel quantum wires. When unpolarized current,
consisting of both spin-$\uparrow$ and spin-$\downarrow$ electrons is injected through
more than one propagating modes, defined by the transverse quantization in  the leads
of a realistic multichannel ring device, each  channel will carry its own phase. Moreover,
during quantum transport involving more than one conducting channel, any spin-independent 
scattering of charge (at interfaces or boundaries  in the case of a  clean system) subjected
to  SO coupling will lead to a loss of coherence of spin quantum state due to the possibility
to entangle spin and orbital degrees of freedom (i.e., within the entangled state, spin
subsystem cannot be described by a pure state spinor).~\cite{decoherence} The multichannel
nature of quantum transport in realistic rings employed in experiments is also part of 
the controversy surrounding recent fundamental pursuits~\cite{shayegan,yang} of the observable
effect of spin Berry phase. The analysis of  spin-dependent features in conductance
oscillations as a function of the magnetic field in disordered rings is not as transparent
as in the case of strictly 1D systems.~\cite{yang} The  AC phase acquired by electrons
propagating through the 1D ring, subjected to the Rashba electric field orthogonal to the
ring plane (Fig.~\ref{fig:ring}), was shown~\cite{su} to consist of a dynamical Rashba phase
(which depends on the cycle duration) and the geometrical Aharonov-Anandan (AA) phase
(which depends only on the path traced in the parameter space) characterizing any
nonadiabatic cyclic evolution.~\cite{geo_phase} In the adiabatic limit, when spin becomes
aligned with the effective momentum-dependent Rashba magnetic field ${\bf B}_{\rm Rashba}({\bf k})$ 
in the reference frame of transported electron (Fig.~\ref{fig:ring}),~\cite{diego}
the AA phase becomes the spin-orbit Berry phase introduced in Ref.~\onlinecite{aronov}.

Here we address the problem of unpolarized current modulation in clean mesoscopic 2D
rings with Rashba SO coupling by studying how the pattern of conductance oscillations
changes as one opens conducting channels, one-by-one, for quantum transport. The paper
is organized as follows. In Sec.~\ref{sec:hamiltonian} we discuss the issue of suitable 
Hamiltonian description of the ring with SO interactions. Then we introduce a model for
the multichannel ring which makes it possible to efficiently implement real$\otimes$spin
space Green function technique in order to obtain both the spin and the charge properties
of {\em coupled spin-charge} transport.~\cite{decoherence} Section~\ref{sec:onech} studies 
the conductance modulation in strictly 1D rings as a function of the Fermi energy $E_F$ of the
zero-temperature quantum transport. We also establish in this section the connection between
the orientation of spin polarization vector and spin precession induced by the Rashba SO interaction, 
pointing out at spin-switch device properties of the AC rings when injected
current is fully polarized. In Sec.~\ref{sec:multich} we discuss conductance modulation in
2D rings with two and three open conducting channels. Although we find that the ring
conductance, in general, is not diminished to zero in the multichannel transport at any
strength of the SO coupling, it still displays oscillations as we tune the strength of
the Rashba interaction. In order to relate such ``incomplete'' conductance modulation to
spin interference effects of states which are not pure (but are instead described by the
density matrices), we analyze in Sec.~\ref{sec:b} transport through 2D rings within the
picture of transmission eigenchannels.~\cite{carlo_rmt} This makes it possible to view
the multichannel transport as if occurring through a set of ``independent'' 1D rings
whose channels, however, can have complicated spin coherence properties. We conclude in 
Sec.~\ref{sec:conclusion}.

\section{Hamiltonian models for ring with Rashba Spin-Orbit interaction} \label{sec:hamiltonian}

In the absence of  external magnetic field, the ballistic semiconductor ring structure
subjected to the Rashba SO coupling is described by the following effective mass
single-particle Hamiltonian
\begin{eqnarray} \label{eq:rashba_hamil}
\hat{H}_{\rm 2D}=\frac{\hat{\vect{p}}^2}{2m^*}
+
\frac{\alpha}{\hbar}\left(\hat{\vectg{\sigma}}\times\hat{\vect{p}}\right)_z
+
V_{\rm conf}(r),
\end{eqnarray}
where $\hat{\vectg{\sigma}}$ is the vector of the Pauli spin operators,
$\hat{\vect{p}}$ is the momentum vector in 2D space, and $\alpha$ is
the strength of the Rashba SO coupling. The last term $V_{\rm conf}(r)$
represents the potential which confines electrons to a finite ring region
within 2DEG.
The analytical expressions for the ring conductance as a function of $\alpha$ have
been obtained only for a strictly 1D ring geometry---in this limit one can find
the eigenstates of a closed ring and then compute the transmission coefficients by
opening the ring to the attached leads where electrons are injected at the Fermi energy
equal to the eigenenergies.~\cite{ring,diego,molnar} However, a search for the correct
1D ring Hamiltonian, which includes the Rashba interaction, has turned out to be an
ambiguous task yielding several apparent solutions. For example, some of the recent
studies have employed non-Hermitian Hamiltonians.~\cite{aronov,ring} It was pointed in
Ref.~\onlinecite{meijer} that performing the limit from 2D to 1D carefully, by taking into
account the radial wavefunctions in the presence of narrow confinement, leads to a
unique Hermitian single-particle Hamiltonian in cylindrical coordinates
\begin{eqnarray} \label{eq:ring_hamil}
\hat{H}_{\rm 1D}(\phi)&=&
\frac{-\hbar^2}{2m^* R^2}\frac{\partial^2}{\partial \phi^2}
-
\frac{i\alpha}{R}\left(\cos\phi\sigma_x+\sin\phi\sigma_y\right)\frac{\partial}{\partial \phi}
\nonumber\\
&&-
\frac{i\alpha}{2R}\left(\cos\phi\sigma_y-\sin\phi\sigma_x\right).
\end{eqnarray}
Here $R$ is the radius of the 1D ring, and $\phi$ is the angular coordinate.
The last term in Eq.~(\ref{eq:ring_hamil}), which disappears if we simply take the radial
coordinate $r=R$ in the cylindrical coordinate representation of the 2D Hamiltonian,
is actually indispensable for this Hamiltonian to be a usual Hermitian operator
corresponding to quantum observable.

\begin{figure}[ht]
\begin{center}
{\includegraphics[scale=.33]{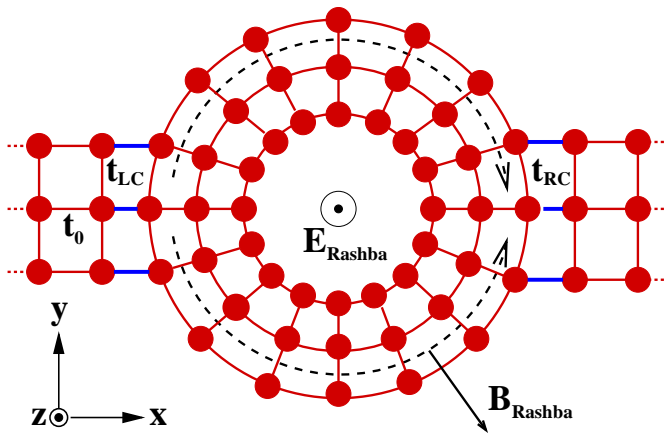}}
\end{center}
\caption{Schematic illustration of the finite-width ring mesoscopic conductor modeled
by a concentric tight-binding lattice, where the values of the parameters $M$ and $N$ in
the Hamiltonian Eq.~(\ref{eq:tight_ring_hamil}) are chosen to be 3 and 16, respectively.
In the actual numerical calculations, we use $M=$1--3 and $N=200$. Here ${\bf E}_{\rm Rashba}$
is the confining electric field for 2DEG (which induces the Rashba SO interaction)
and ${\bf B}_{\rm Rashba}$ is the corresponding effective (momentum-dependent) magnetic
field in the reference frame of transported spin.}
\label{fig:ring}
\end{figure}

While the 1D Hamiltonian Eq.~(\ref{eq:ring_hamil}) provides a simplest starting point to study
quantum transport through the ring within the Landauer-B\" utikker transmission formalism,~\cite{meso} 
for efficient and {\em numerically exact} treatment of multichannel transport through finite width 2D rings 
attached to the leads it is advantageous to switch to a local orbital basis representation that makes is 
possible to employ real$\otimes$spin space Green function technique and obtain.~\cite{decoherence} 
Therefore, we introduce here a concentric tight-binding lattice Hamiltonian, composed of  $M=1,2, \ldots$ 
concentrically connected tight-binding ring chains, as sketched in Fig.~\ref{fig:ring}.
The two ideal semi-infinite leads with $\alpha=0$,
of the same width as the ring itself, are attached symmetrically to the ring thereby breaking
the rotational invariance of the closed ring problem. The $M=1$ case corresponding to a single ring chain, 
which represent a lattice version of the correct 1D Hamiltonian in  Eq.~(\ref{eq:ring_hamil}),
has frequently been employed to study the Aharonov-Bohm effect
in mesoscopic ring-shaped conductors.~\cite{abchain} The $M=2$ case has appeared in the studies of 
the influence of finite ring width on the Aharonov-Bohm oscillations of magneto-conductance.~\cite{aldea} 
Here we introduce the Rashba SO interaction into the
concentric tight-binding lattice with arbitrary number $M$ of ring chains.  
One of the  advantages of our model over the conventional square lattice discretization of the finite
width rings~\cite{abring} is that the width of the ring can be controlled precisely by
varying the number of the ring chains one-by-one. This makes it possible to study the effects
of multichannel transport systematically. The maximum number of open channels in a structure 
consisting of $M$ ring chains is equal to $M$.
The Hamiltonian corresponding to the set-up depicted in Fig.~\ref{fig:ring} contains five
terms
\begin{eqnarray} \label{eq:five_terms}
\hat{H}=\hat{H}_{\rm ring}+\hat{H}_{\rm L}+\hat{H}_{\rm R}+\hat{V}_{\rm L,ring}+\hat{V}_{\rm R,ring}.
\end{eqnarray}
The first term, which describes electrons in an isolated (i.e., closed to the environment)
ring that are subjected to the Rashba SO coupling, is given by
\begin{widetext}
\begin{eqnarray}\label{eq:tight_ring_hamil}
\hat{H}_{\rm ring}
&=&
\sum_{n=1}^{N}
\sum_{m=1}^{M}
\sum_{\sigma=\up,\dn}
\ep_{nm} \hat{c}^\dagger_{nm;\sigma}\hat{c}_{nm;\sigma}
-
\sum_{n=1}^{N}
\sum_{m=1}^{M}
\sum_{\sigma,\sigma'=\up,\dn}
\Biggl[
t_{\phi}^{n,n+1,m;\sigma,\sigma'}
\hat{c}^\dagger_{nm;\sigma}\hat{c}_{n+1,m;\sigma'}
+
{\rm h.c.}
\Biggr]
\nonumber\\
&&
-\sum_{n=1}^{N}
\sum_{m=1}^{M-1}
\sum_{\sigma,\sigma'=\up,\dn}
\Biggl[
t^{m,m+1,n;\sigma,\sigma'}_{r}
\hat{c}^\dagger_{nm;\sigma}\hat{c}_{n,m+1;\sigma'}
+
{\rm h.c.}
\Biggr].
\end{eqnarray}
\end{widetext}
Here $n=1,2, \ldots, N$ and $m =1,2,\ldots,M$ are the lattice site indices along
the azimuthal ($\phi$) and the radial ($r$) directions, respectively.
The operator $c_{nm;\sigma}$ ($c^\dagger_{nm;\sigma}$)
annihilates (creates) a spin $\sigma$ electron at the site
$(n,m)$ of the ring. In our notation $m=1$ corresponds to the innermost ring
chain, while $m=M$ stands for the outermost ring chain to which the external
leads are attached. The operator $c_{N+1,m;\sigma}$ ($c_{N+1,m;\sigma}^{\dagger}$)
is identified with $c_{1,m;\sigma}$ ($c_{1,m;\sigma}^{\dagger}$)
due to the periodic boundary condition. In Eq.~(\ref{eq:five_terms}) $\ep_{nm}$ is the
on-site potential, while $t_{\phi}^{n,n+1,m;\sigma,\sigma'}$ and $t^{m,m+1,n;\sigma,\sigma'}_{r}$ 
are the nearest neighbor hopping energies along the radial and the angular directions, respectively. 
Those hopping energies have been generalized to include the SO coupling
terms, which are given in the $2\times 2$ matrix form as
\begin{eqnarray}\label{eq:hopping}
t_{\phi}^{n,n+1,m} & = &
\frac{1}{(r_m/a)^2 \Delta\phi^2}t_0 \hat{I}_s \nonumber \\
&& -i\frac{t_{\rm so}}{(r_m/a)\Delta\phi}(\cos\phi_{n,n+1}\sigma_x+
\sin\phi_{n,n+1}\sigma_y), \nonumber \\
t^{m,m+1,n}_{r} &=&
t_0 \hat{I}_s + it_{\rm so}(\cos\phi_n\sigma_y-\sin\phi_n\sigma_x).
\end{eqnarray}
Here $\phi_{n}\equiv 2\pi(n-1)/N$, $\phi_{n,n+1}\equiv
(\phi_n+\phi_{n+1})/2$,
$\Delta\phi\equiv 2\pi/N$,
$r_m\equiv r_{1}+(m-1)a$, $t_0\equiv \hbar^2/2ma^2$ with $a$ being the
lattice spacing constant along the radial direction,
$t_{\rm so} \equiv \alpha/2a$ is the tight-binding
SO coupling energy with $\alpha$ being the SO coupling strength of
the original Hamiltonian Eq.~(\ref{eq:rashba_hamil}),
and $\hat{I}_s$ is the $2\times 2$identity matrix. We further assume that the
lattice spacing along the azimuthal direction in the outermost ring
chain ($m=M$) is the same as that of the radial direction, such that
$r_M\Delta\phi(=2\pi r_M/N)\equiv a$. In order to avoid the negative value of
$r_1$ (the radius of the innermost chain), the value of $M$ has to satisfy
the condition $M<N/2\pi+1$ .

In Eq.~(\ref{eq:five_terms}), the second term is
the Hamiltonian for the left lead
\begin{eqnarray} \label{eq:left_lead}
\hat{H}_{\rm L}=
-t_0
\sum_{i=1}^{\infty}
\sum_{j=1}^{M-1}
\sum_{\sigma=\up,\dn}
\Biggl[
\hat{b}^\dagger_{{\rm L},i,j;\sigma}
\hat{b}_{{\rm L},i+1,j,\sigma}
+{\rm h.c.}
\Biggr],
\end{eqnarray}
where $i$ and $j$ are the lattice indices along the $x$ (current flowing)
and the $y$ (transverse) directions, respectively. The operators $b_{{\rm L},i,j,\sigma}$ 
($b^{\dagger}_{{\rm L},i,j,\sigma}$) annihilate (create) spin $\sigma$ electron at the site
$(i,j)$ in the left lead. The Hamiltonian of the right lead $\hat{H}_{\rm R}$ has the same
form as Eq.~(\ref{eq:left_lead}). Finally, the coupling between the leads and ring is
described by the following Hamiltonians
\begin{eqnarray}\label{eq:coupling}
\hat{V}_{{\rm L,ring}}&=&-
t_{{\rm LC}}
\sum_{k=1}^{M}
\sum_{\sigma=\up,\dn}
\Biggl[
\hat{b}^\dagger_{{\rm L},1,k;\sigma}\hat{c}_{k,M;\sigma}
+{\rm h.c.}
\Biggr],
\\
\hat{V}_{{\rm R,ring}}&=&-
t_{{\rm RC}}
\sum_{k=1}^{M}
\sum_{\sigma=\up,\dn}
\Biggl[
\hat{b}^\dagger_{{\rm R},1,k;\sigma}\hat{c}_{\frac{N}{2}+k,M;\sigma}
+{\rm h.c.}
\Biggr].
\end{eqnarray}
Here $t_{{\rm L(R)}C}$ is the coupling strength between the left (right) lead and the ring.
It is assumed that the left and the right lead are attached to the
ring symmetrically, where we neglect the finite curvature of the outermost ring
at the interface between the ring and the leads (this is justified if when the 
condition $N \gg M$ is satisfied).

Once the Hamiltonian Eq.~(\ref{eq:five_terms}) is given, one can evaluate the
matrix of spin-resolved conductance by using the Landauer-B\"uttiker's formula
generalized to include the spin-degree of freedom
\begin{eqnarray} \label{eq:conductance}
{\bf G}=
\left(
\begin{array}{cc}
G^{\up\up} & G^{\up\dn}
\\
G^{\dn\up} & G^{\dn\dn}
\end{array}
\right)
=
\frac{e^2}{h}
\sum_{p,p^\prime=1}^M
\left(
\begin{array}{cc}
\left|{\bf t}_{p^\prime p,\up\up}\right|^2 & \left|{\bf t}_{p^\prime p,\up\dn}\right|^2
\\
\left|{\bf t}_{p^\prime p,\dn\up}\right|^2 & \left|{\bf t}_{p^\prime p,\dn\dn}\right|^2
\end{array}
\right),
\nonumber\\
\end{eqnarray}
where ${\bf t}_{p^\prime p,\sigma^\prime \sigma}$ is a transmission matrix element
determining the probability amplitude that electron injected in orbital channel
$|p \rangle$ with spin $\sigma$ in the left lead would end up in a conducting channel 
$|p^\prime \rangle$ of the right lead with spin $\sigma^\prime$ (the index $p=1,2,\ldots$ labels 
the quantized transverse propagating modes in the leads, while $\sigma=\up,\dn$ describes the
spin state). The transmission matrix is calculated using the real$\otimes$spin space 
Green function technique.~\cite{decoherence} Note that the spin-quantization axis can be chosen arbitrarily. 
For example, for the spin-quantization axis chosen along the $x$-direction, $G^{\dn\up}$ can be 
interpreted as the conductance for a two-probe set-up where the electrodes are half-metallic 
ferromagnets---magnetization of the left lead is parallel while the magnetization of the right lead 
is antiparallel to the $x$-axis. The {\em total} conductance characterizing conventional  unpolarized charge transport
\begin{equation} \label{eq:total}
G^{\rm tot}= G^{\up\up} + G^{\up\dn} + G^{\dn\up} + G^{\dn\dn}
\end{equation}
is independent of the arbitrariliy chosen spin-quantization axis for the calculation of spin-resolved transport properties.

If a quantum system is fully coherent, its state is described by a density matrix $\hat{\rho}^2=\hat{\rho}$.  The decrease of the degree of quantum coherence due to
entanglement to environment ({\em decoherence})  or other dephasing processes (such
as classical noise) can be quantified by the purity~\cite{zurek} $\zeta = {\rm Tr} \,
\hat{\rho}^2$. Since in the case of the spin-$\frac{1}{2}$ particle
$\zeta_s = (1+|{\vect{P}}|^2)/2$ depends solely  on the modulus of the spin-polarization
(Bloch) vector $\vect{P}$
\begin{equation} \label{eq:density_matrix}
\hat{\rho}_s=\left( \begin{array}{cc}
        \rho_{\uparrow\uparrow} & \rho_{\uparrow\downarrow} \\
        \rho_{\downarrow\uparrow} & \rho_{\downarrow\downarrow}
         \end{array} \right)= \frac{\hat{I}_s+ {\vect{P}} \cdot \hat{\bm{\sigma}}}{2},
\end{equation}
$|\vect{P}|$ can be used to measure the degree of coherence retained in spin states in
the course of their transport through complicated semiconductor environment.~\cite{decoherence}

While the conductance formula Eq.~(\ref{eq:conductance}) requires to evaluate only the
amplitude of the complex transmission matrix element ${\bf t}_{p^\prime p,\sigma^\prime \sigma}$,
the evaluation of the spin-polarization vector $\vect{P}$ requires not only the amplitude
but also the phase of the transmission matrix elements. Suppose that we inject 100\%
spin-$\sigma$ polarized current from the left lead. Then the density matrix of the spin
degree of freedom for the outgoing current is given by~\cite{decoherence}
\begin{eqnarray}\label{eq:rho_s}
\hat{\rho}^{\sigma}
&=&
\frac{e^2/h}{G^{\up\sigma}+G^{\dn\sigma}}
\sum_{p,p^\prime=1}^M
\left(
\begin{array}{cc}
\left|{\bf t}_{p^\prime p,\up\sigma}\right|^2 & {\bf t}_{p^\prime p,\up\sigma}t^*_{p^\prime p,\dn\sigma}
\\
{\bf t}_{p^\prime p,\dn \sigma} {\bf t}^*_{p^\prime p,\up\sigma} & 
\left|{\bf t}_{p^\prime p,\dn\sigma}\right|^2
\end{array}
\right).
\nonumber\\
\end{eqnarray}
The measurement of any observable quantity on the spin subsystem in the right lead is
evaluated using such spin density matrix. An example is the {\em current spin polarization
vector}, which is obtained as the expectation value the spin operator $\hat{\vectg{\sigma}}$
\begin{eqnarray}\label{eq:Pvector}
\vect{P}^{\sigma}=
{\rm Tr}_{\rm s}\left[
\hat{\rho}_{}^{\sigma}
\hat{\vectg{\sigma}}
\right],
\end{eqnarray}
where ${\rm Tr}_{\rm s}$ is the trace in the spin Hilbert space.
For example, if the injected current is spin-$\up$ polarized along
\begin{figure}[ht]
\begin{center}
\vspace*{-0.5cm}
{\includegraphics[scale=.3]{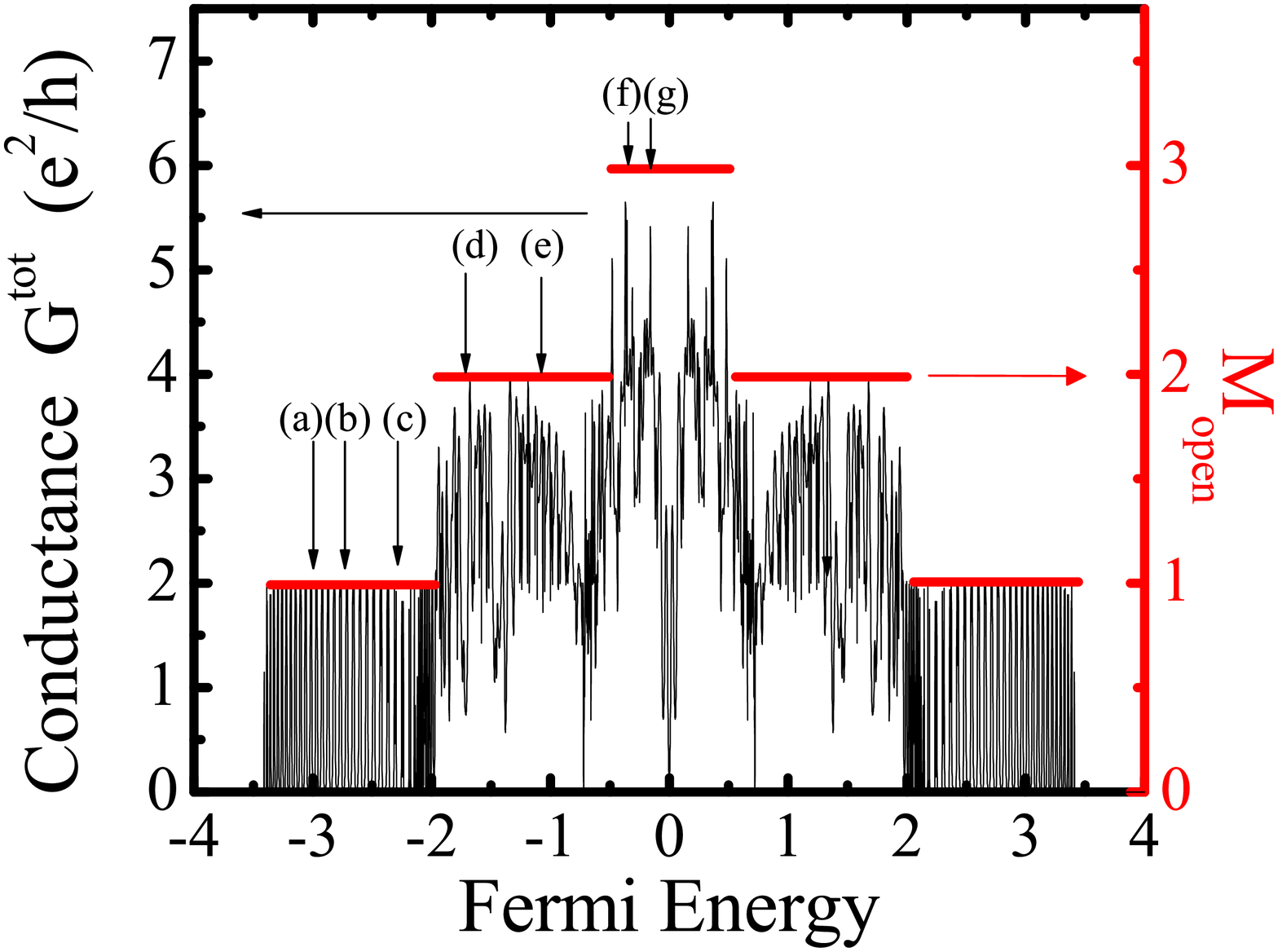}}
\vspace*{-1cm}
\end{center}
\caption{Fermi energy dependence of the total conductance
$G^{\rm tot}=G^{\up\up}+G^{\dn\up}+G^{\up\dn}+G^{\dn\dn}$
of a finite width AC ring $(M,N)=(3,200)$, for a chosen
Rashba SO coupling strength $Q_{\rm R}=6$ [$Q_{\rm R} \equiv t_{\rm so}N /t_0 \pi$].
The red line shows the number of open (orbital) conducting channels
$M_{\rm open}$ at $E_F$. The vertical arrows label the values of the Fermi energy
selected for Fig.~\ref{fig:2d_single_total} (Sec.~\ref{sec:onech}) and Fig.~\ref{fig:2d_multi_total} (Sec.~\ref{sec:multich}).
}
\label{fig:2d_multi_energy}
\end{figure}
the $x$-direction, the spin polarization vector $\vect{P}^{\up}=(P_x^{\up},P_y^{\up},P_z^{\up})$ in the
right lead is given by
\begin{eqnarray}\label{eq:Pvector_elmts}
P^{\up}_x &  = &  \frac{G^{\up \up} -
G^{\dn \up}}{G^{\up \up} + G^{\dn \up}}, \\
P^{\up}_y & = & \frac{2e^2/h}{G^{\up \up} + G^{\dn\up}}
\sum_{p=1}^{M} {\rm Re}
\left[{\bf t}_{p'p, \up \up} {\bf t}^*_{p'p,\dn \up} \right], \\
P^{\up}_z & = & \frac{2e^2/h}{G^{\up \up} + G^{\dn\up}}
\sum_{p'p=1}^{M} {\rm Im}
\left[{\bf t}_{p'p, \up \up}
{\bf t}^*_{p' p,\dn \up} \right],
\end{eqnarray}
where $G^{\up\up}$ and $G^{\dn\up}$ are the spin conserved and the spin flipped
conductance matrix elements. The $x$-axis is chosen arbitrarily as the spin-quantization
axis, $\hat{\sigma}_x = + |\!\! \uparrow \rangle$ and $\hat{\sigma}_x =  - |\!\! \downarrow \rangle$, 
so that Pauli spin algebra has the following representation
\begin{equation}\label{eq:pauli}
 \hat{\sigma}_x=  \left( \begin{array}{cc}
       1 & 0 \\
       0 & -1
  \end{array} \right),
   \hat{\sigma}_y = \left( \begin{array}{cc}
       0 & 1 \\
       1 & 0
  \end{array} \right),
    \hat{\sigma}_z = \left( \begin{array}{cc}
       0 & i \\
       -i & 0
  \end{array} \right).
\end{equation}
specifies the particular form of the expectation values of $\vect{P}$.
One can obtain analogous expressions for $(P_x,P_y,P_z)$  when injected current is
polarized along other directions, as well as for the injection of partially
polarized current.~\cite{decoherence} Thus, studying $\vect{P}$, in addition
to the ring conductance, will allow us to understand spin orientation and degree of
spin coherence that corresponds to modulation of the charge current.

As an example of quantum transport properties of our model Hamiltonian
Eq.~(\ref{eq:five_terms}), Fig.~\ref{fig:2d_multi_energy} plots the total
conductance of the finite width ring $(M,N)=(3,200)$ as a function of $E_F$.
The strength of the Rashba SO coupling is fixed at $Q_{\rm R}=6$, where we introduce
the dimensionless Rashba SO parameter $Q_{\rm R}\equiv 2m \alpha r_M/\hbar^2=(t_{\rm so}/t_0)N/\pi$ with $r_{M}$ as the radius of the outermost ring. Furthermore, we assume that the system is free from impurities $\ep_{nm}=0$, and that coupling energies between the conductor and the leads are set to be the same as the hopping energy in the leads
$t_{\rm LC}=t_{\rm RC}=t_0$. The calculated conductance $G^{\rm tot}(Q_R=6,E_F)$ exhibits
rapid oscillations with peaks occurring at eigenenergies of the corresponding closed 2D
ring. While the conductance oscillates regularly in the single-channel regime, the
oscillation pattern in the multichannel regime is rather intricate since the occupation
of the higher radial modes gives rise to new conductance peaks in addition to the ones originating from the lowest radial mode.

\begin{figure}[ht]
\begin{center}
\vspace*{-0.5cm}
{\includegraphics[scale=.35]{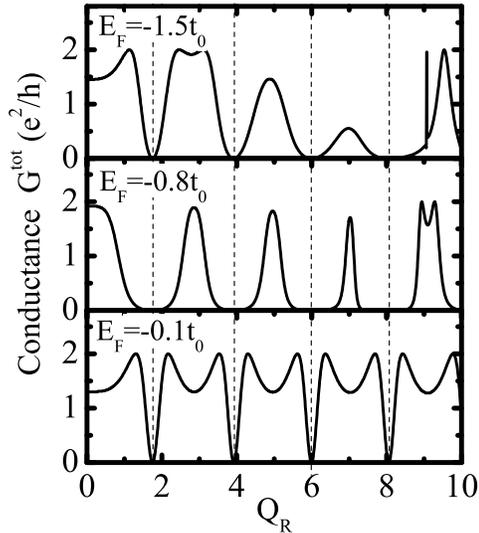}}
\vspace*{-0.5cm}
\end{center}
\caption{The total conductance $G^{\rm tot}(Q_R,E_F)$ of a strictly 1D ring $(M,N)=(1,200)$,  attached to two 1D semi-infinite ideal leads, as a function of the Rashba SO interaction
$Q_{\rm R}$, for three different values of the Fermi energy as a parameter. The vertical
dotted lines denote the position of the conductance minima $G^{\rm tot} \simeq 0$, which do
not depend on $E_F$ at which the zero-temperature quantum transport takes place.}
\label{fig:1d_total}
\end{figure}
\begin{figure}[ht]
\begin{center}
\vspace*{-0.5cm}
{\includegraphics[scale=.25]{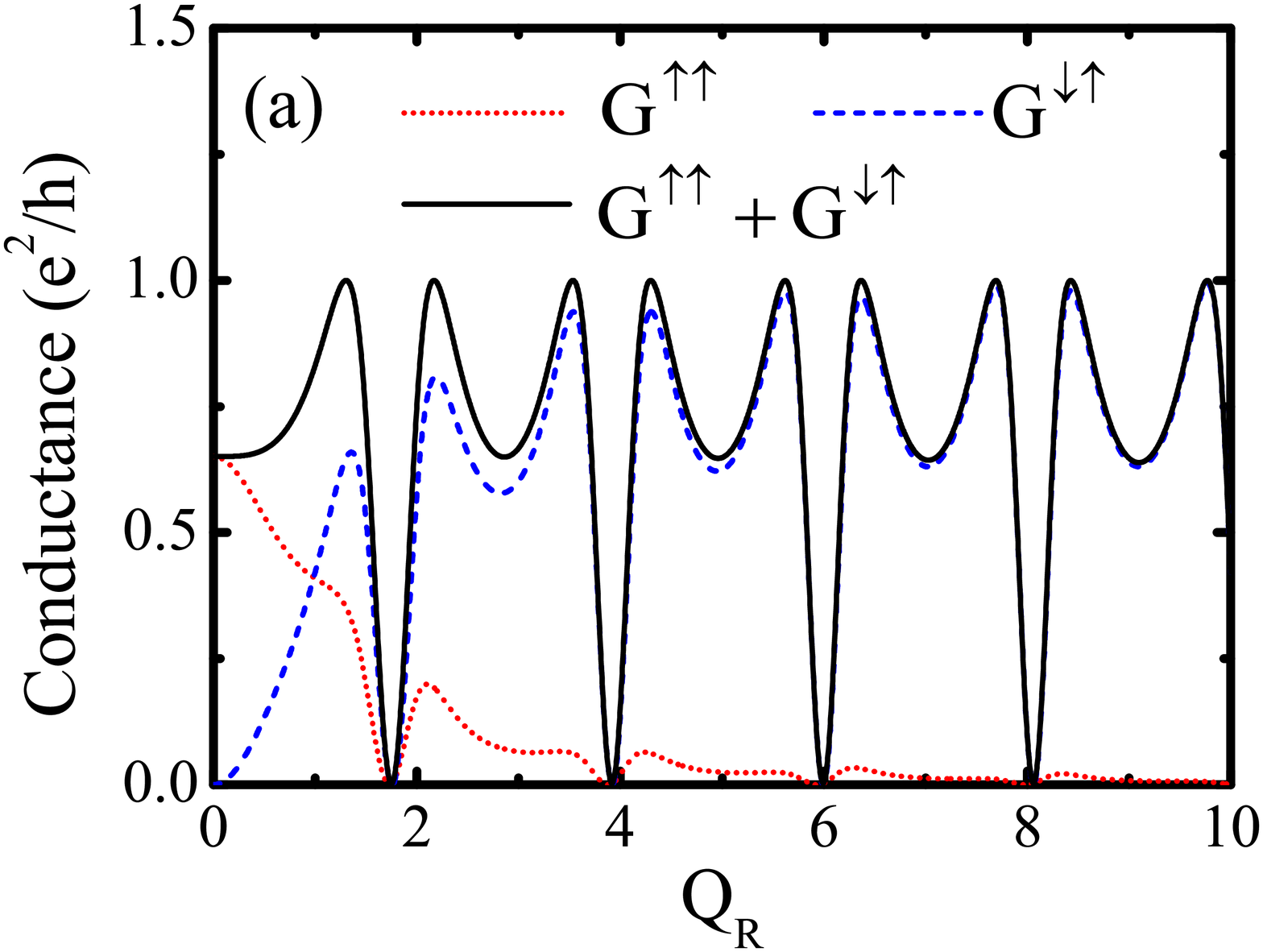}}
\vspace*{-0.5cm}
\end{center}
\begin{center}
\vspace*{-0.5cm}
\hspace*{0.05cm}
{\includegraphics[scale=.25]{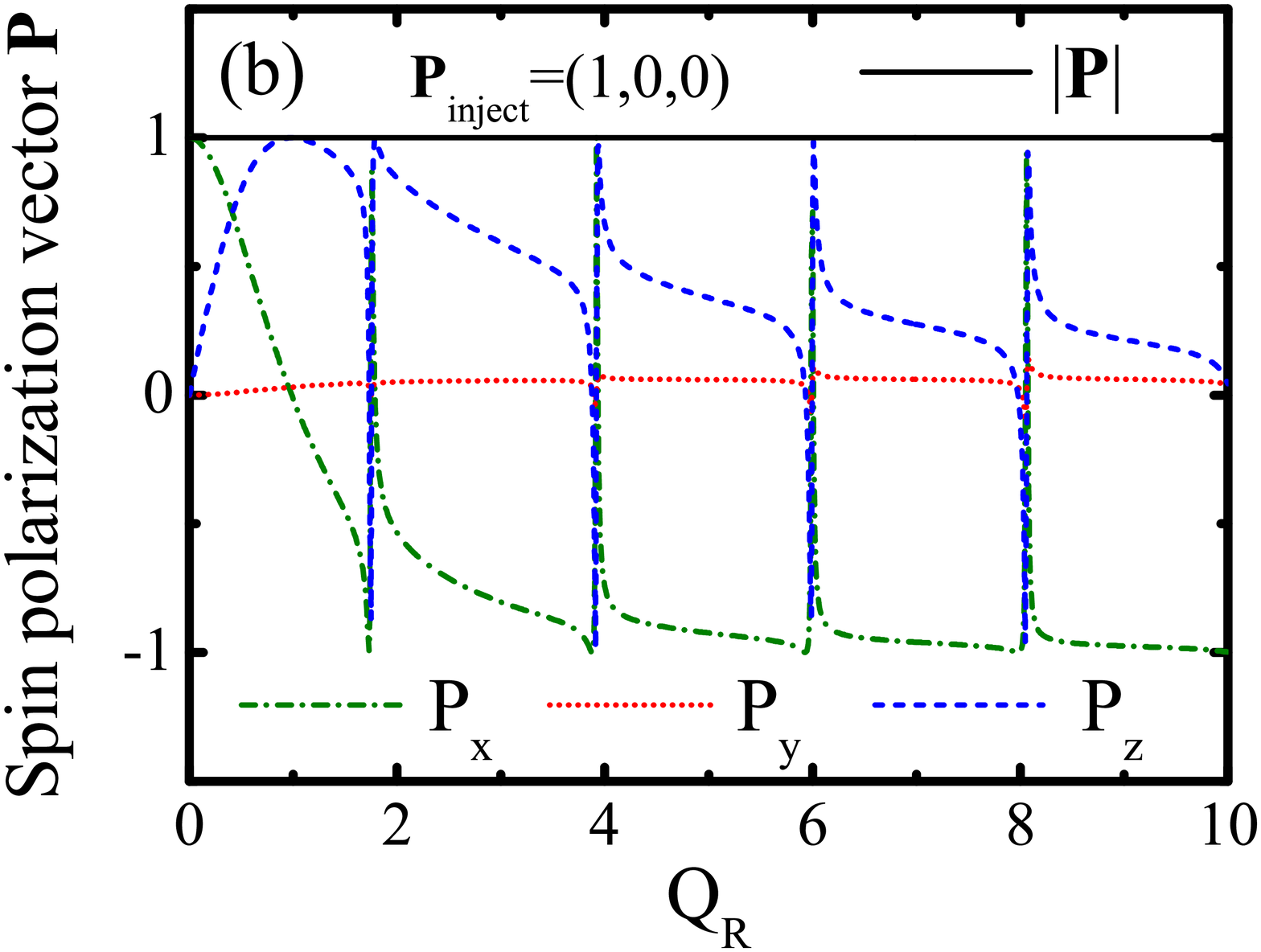}}
\vspace*{-0.5cm}
\end{center}
\caption{The spin-resolved conductances (a) $G^{\sigma,\sigma^\prime}(Q_R,E_F)$
and the outgoing current spin polarization vector (b) $\vect{P}(Q_R,E_F)$ versus
the Rashba SO interaction $Q_{\rm R}$ for a strictly 1D ring conductor with
$(M,N)=(1,200)$ and Fermi energy $E_F=-0.1t_0$. The spin-quantization axis is
chosen to be the $x$-axis.
}
\label{fig:1d_spin}
\end{figure}
\begin{figure}[ht]
\begin{center}
\vspace*{-0.5cm}
{\includegraphics[scale=.35]{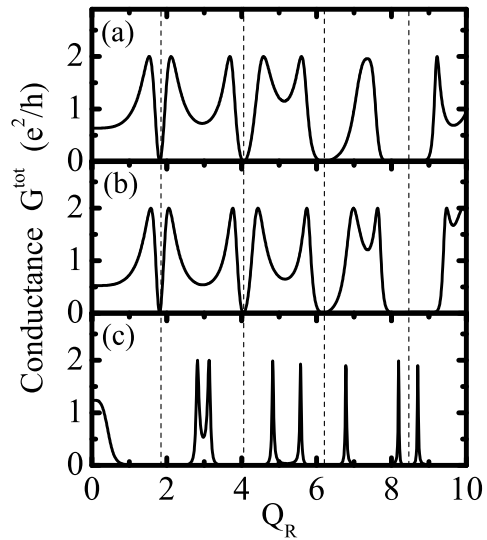}}
\vspace*{-0.5cm}
\end{center}
\caption{The total conductance $G^{\rm tot}$ vs. the Rashba SO interaction $Q_{\rm R}$
of a finite-width ring conductor with $(M,N)=(3,200)$ for three different values of
the Fermi energy: (a) $E_F=-3.0t_0$, (b) $E_F=-2.7t_0$, and (c) $E_F=-2.2t_0$, which
allow only one channel to propagate (see Fig.~\ref{fig:2d_multi_energy}). The vertical
dotted lines indicate the minima of the conductance $G^{\rm tot} \simeq 0$, which do not
depend on $E_F$.}
\label{fig:2d_single_total}
\end{figure}
\begin{figure}[ht]
\begin{center}
\vspace*{-0.5cm}
{\includegraphics[scale=.35]{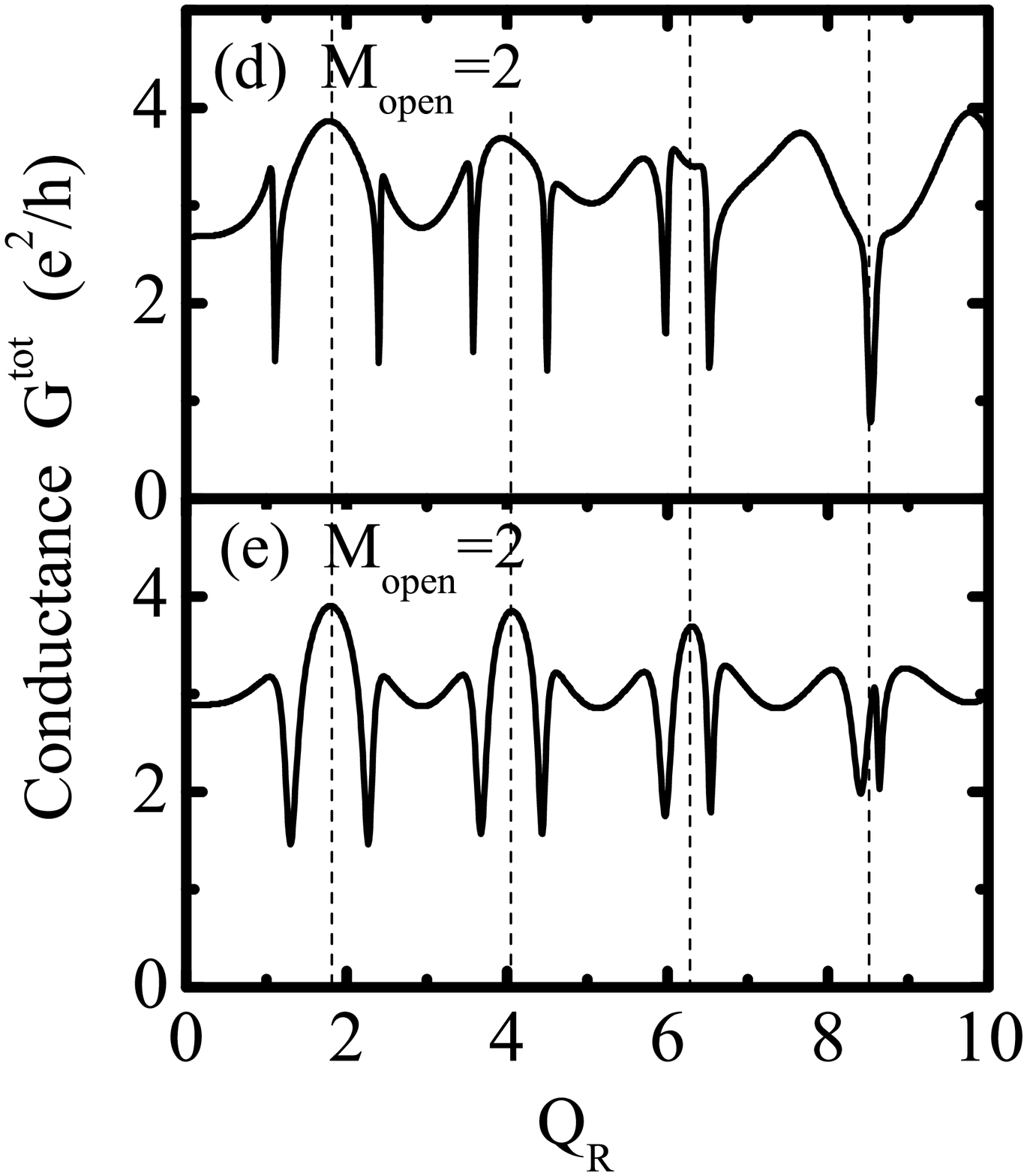}}
\vspace*{-0.5cm}
\end{center}
\begin{center}
\vspace*{-0.5cm}
{\includegraphics[scale=.35]{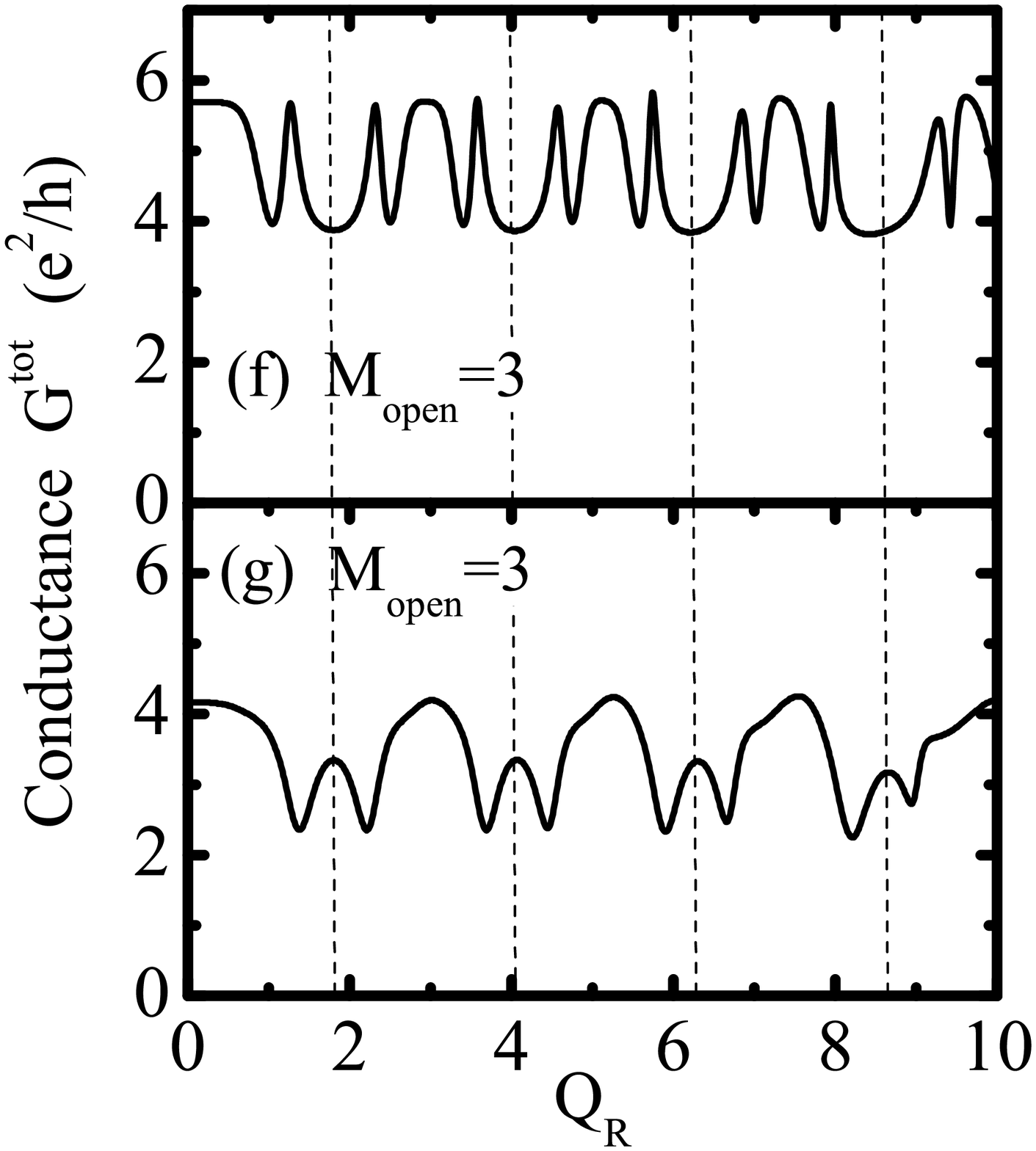}}
\vspace*{-0.5cm}
\end{center}
\caption{
The total conductance $G^{\rm tot}$ vs. the Rashba SO interaction $Q_{\rm R}$ (upper panel)
for a finite-width ring conductor $(M,N)=(3,200)$ at two different values of the Fermi
energy: (d) $E_F=-1.8t_0$ and (e) $E_F=-1.0t_0$ at which the number of open conducting
channels is $M_{\rm open}=2$. In the lower panel the selected Fermi energies, (f) $E_F=-0.35t_0$
and (g) $E_F=-0.1t_0$, determine $M_{\rm open}=3$ (see arrows in Fig.~\ref{fig:2d_multi_energy}).
}
\label{fig:2d_multi_total}
\end{figure}
\begin{figure}[ht]
\begin{center}
\vspace*{-0.5cm}
{\includegraphics[scale=.25]{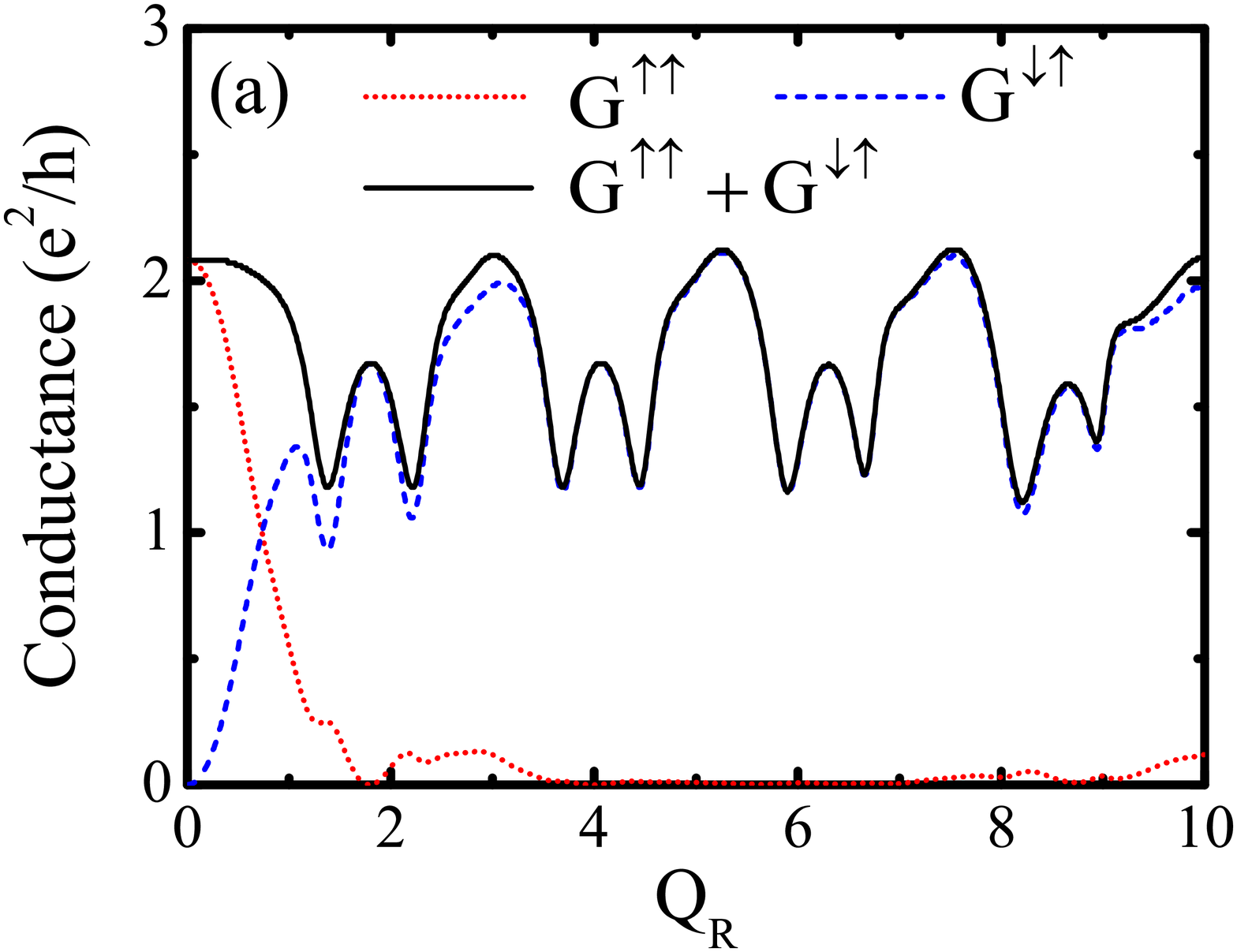}}
\vspace*{-0.5cm}
\end{center}
\begin{center}
\vspace*{-0.5cm}
{\includegraphics[scale=.25]{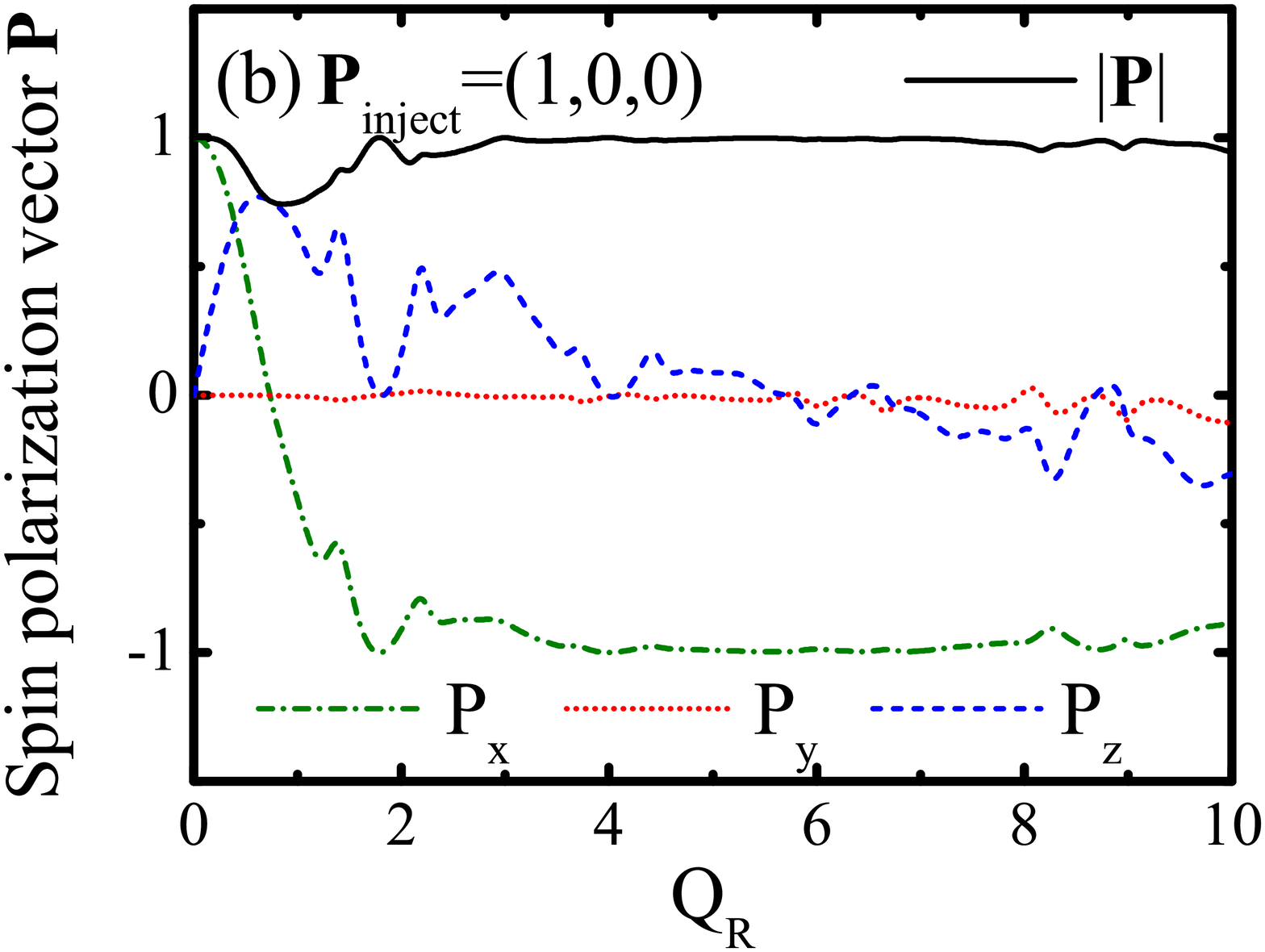}}
\vspace*{-0.5cm}
\end{center}
\caption{The spin-resolved conductances (a) $G^{\sigma,\sigma^\prime}(Q_R,E_F)$
and the outgoing current spin polarization vector (b) $\vect{P}(Q_R,E_F)$ versus
the Rashba SO interaction $Q_{\rm R}$ for a strictly 1D ring conductor with
$(M,N)=(3,200)$ and Fermi energy $E_F=-0.1t_0$. The spin-quantization axis is
chosen to be the $x$-axis.}
\label{fig:2d_multi_spin_1}
\end{figure}
\begin{figure}[ht]
\begin{center}
\vspace*{-0.5cm}
{\includegraphics[scale=.25]{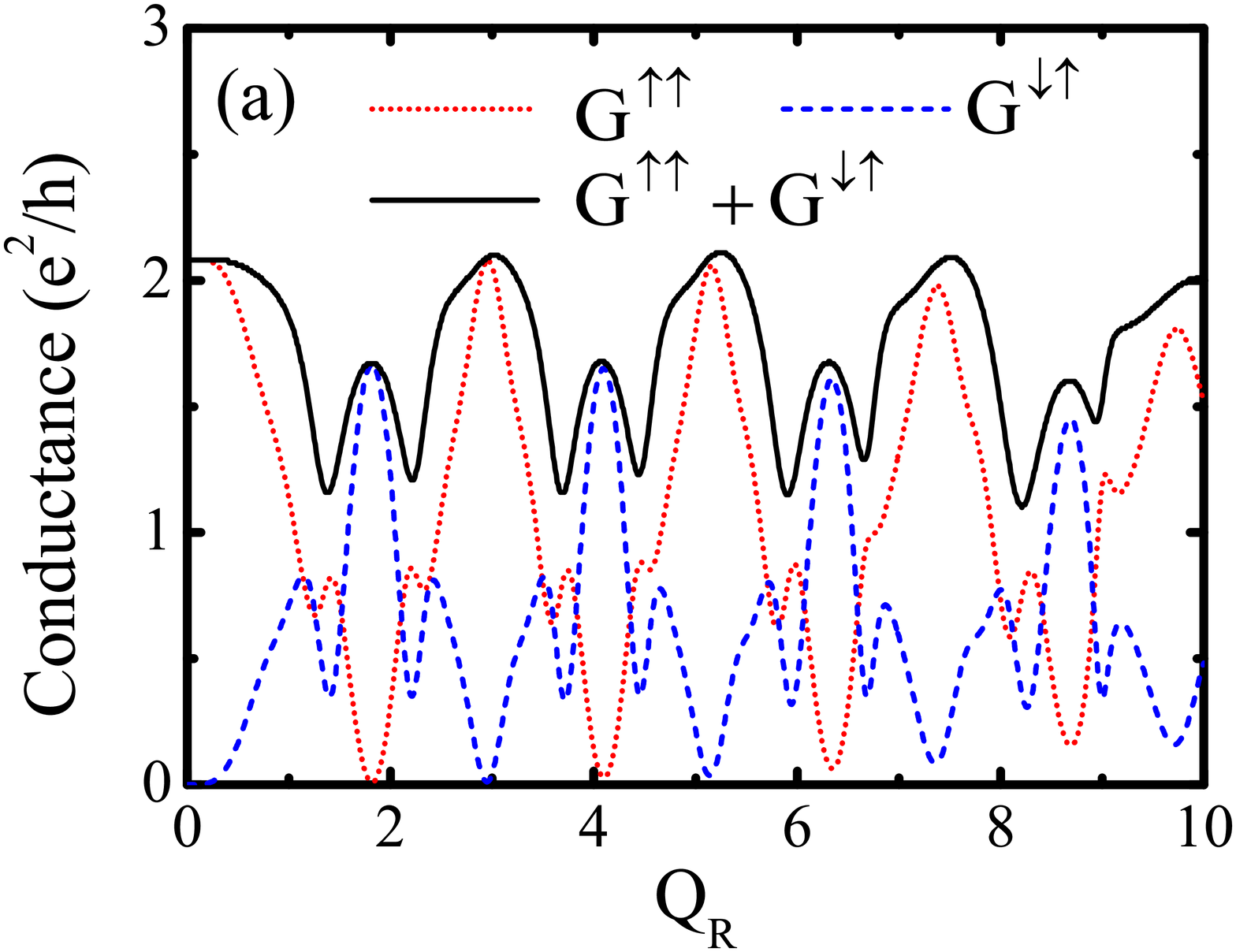}}
\vspace*{-0.5cm}
\end{center}
\begin{center}
\vspace*{-0.5cm}
\hspace*{0.05cm}
{\includegraphics[scale=.25]{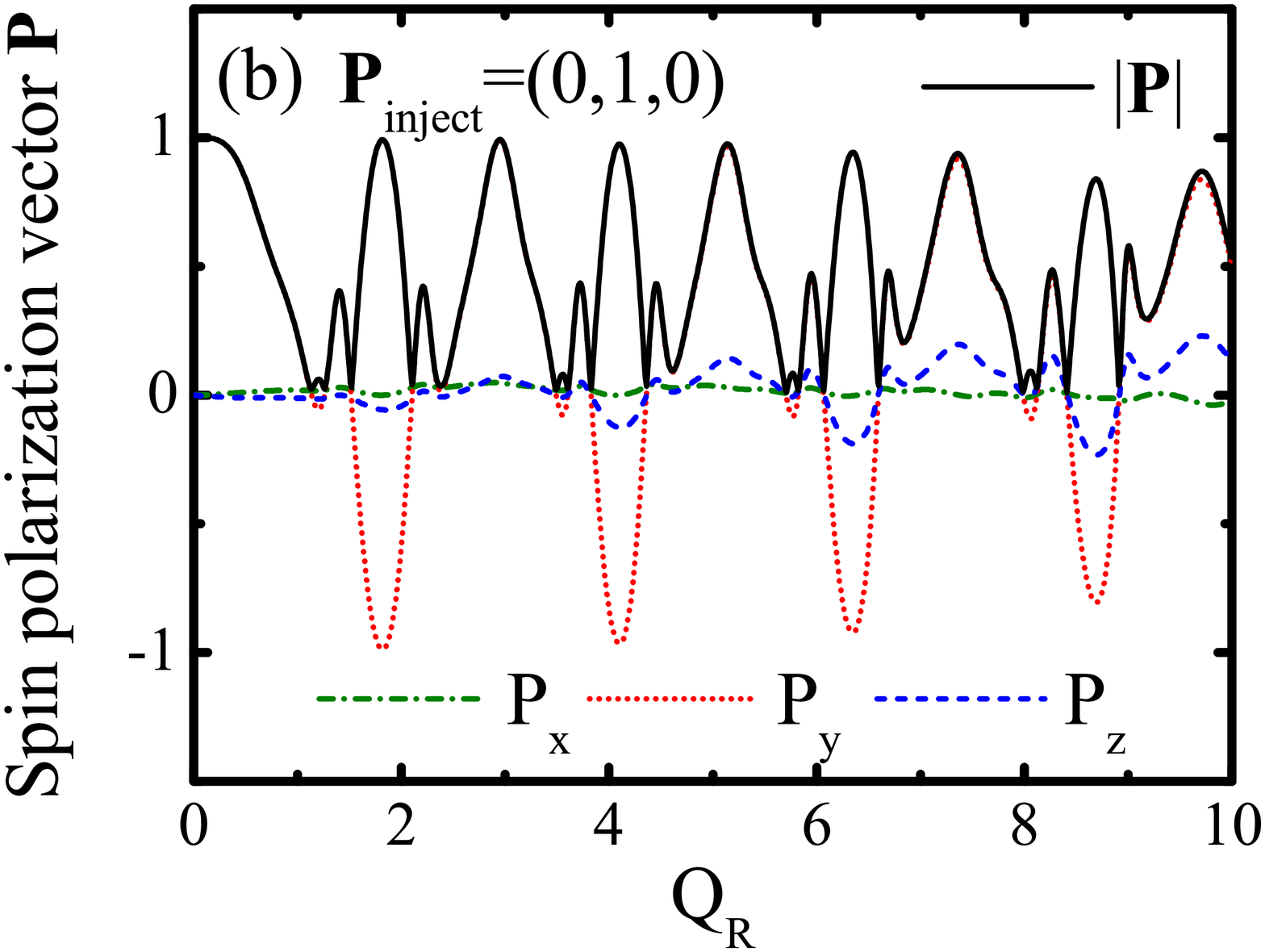}}
\vspace*{-0.5cm}
\end{center}
\caption{The spin-resolved conductances (a) $G^{\sigma,\sigma^\prime}(Q_R,E_F)$
and the outgoing current spin polarization vector (b) $\vect{P}(Q_R,E_F)$ versus
the Rashba SO interaction $Q_{\rm R}$ for a strictly 1D ring conductor with
$(M,N)=(3,200)$ and Fermi energy $E_F=-0.1t_0$. The spin-quantization axis is
chosen to be the $y$-axis.}
\label{fig:2d_multi_spin_2}
\end{figure}
\section{Spin-interference effects in single-channel quantum transport through AC rings} \label{sec:onech}
In this section we study how $G^{\rm tot}$ can be  modulated as we change the strength
of the Rashba SO coupling $Q_R$ in both strictly 1D rings and single open channel of
2D rings. The transport through phase-coherent 1D rings, described by the correct
Hamiltonian~\cite{meijer} Eq.~(\ref{eq:ring_hamil}) has been analyzed recently,~\cite{diego}
in terms of the expressions for $G^{\rm tot}$ of the ring of radius $R$ that does not
involve dependence on the Fermi energy
\begin{eqnarray}\label{eq:analytical}
G^{\rm tot}=e^2/h\left\{1+\cos\left[\pi\left(\sqrt{Q_\alpha^2+1}-
1\right)\right]\right\}.
\end{eqnarray}
Here $Q_\alpha=2m \alpha R/\hbar^2$ has the meaning of the spin precession angle
over the circumference of 1D ring. However, such simplified expression neglects the
back scattering at the interface between the ring and the leads. A more involved
treatment that takes into account such effects has been undertaken in
Ref.~\onlinecite{molnar}. Nevertheless, no analytical expression has been
obtained for single-channel transport in 2D rings.

We confirm~\cite{molnar} in Figure~\ref{fig:1d_total} that the exact pattern of zero-temperature
$G^{\rm tot}$ versus Rashba SO coupling strength depends on the Fermi energy of
transported electrons. This is due to the back scattering effects at the interfaces between
the ring and the leads, which can strongly affect the transport. Nevertheless, all of the calculated
conductance curves have dips $G^{\rm tot}=0$  at specific values of $Q_R$ that are spaced
quasi-periodically~\cite{diego} in a way which does not depend on $E_F$. Thus, the
zeros of the conductance in Fig.~\ref{fig:1d_total} agree well with those predicted
by Eq.~(\ref{eq:analytical}), i.e., their position is insensitive to the lead-ring back
scattering effects.

In order to understand the origin of the conductance modulation in more detail, we plot
the spin-resolved conductances for a given Fermi energy $E_F=-0.1t_0$ in Fig.~\ref{fig:1d_spin}.
Here we consider
the conductances corresponding to the injection of current which is fully spin-$\up$ polarized along
the $x$-axis. As $Q_{\rm R}$ increases, $G^{\up\up}$ decays while the spin-flipped conductance
$G^{\dn\up}$ increases. That is, at the interface between the ring and the left lead,
the injected spin-$\up$ (along the $x$-axis) finds itself to be parallel or antiparallel
to the effective $\vect{k}$-dependent Rashba magnetic field $\vect{B}_{\rm Rashba}(\vect{k})$
that appears in the frame of  electrons circulating along the ring (see Fig.~\ref{fig:ring}).
In the presence of large enough Rashba SO coupling, the injected spin-$\up$
polarized current will change its spin polarization by following the direction of
$\vect{B}_{\rm Rashba}(\vect{k})$ adiabatically. The corresponding outgoing current will
appear in the right lead as spin-$\dn$ polarized current. The summation of those two
components $G^{\up\up}+G~{\dn\up}$ is equal to half of the total conductance
$G^{\rm tot}$ plotted in Fig.~\ref{fig:1d_total}.

The spin polarization vector $\vect{P}=(P_x,P_y,P_z)$ of the transmitted
current in Fig.~\ref{fig:1d_spin} further clarifies this insight. We see that
with increasing SO coupling, $\vect{P}$ in the right lead rotates from
$\vect{P}=\vect{P}_{\rm inject}=(1,0,0)$ at $Q_R \rightarrow 0$ to the asymptotic
value $\vect{P} \approx (-1,0,0)$ at large $Q_R$. The functions $P_x(Q_R)$ and $P_z(Q_R)$
are, however, not monotonous since there are abrupt changes of their directions at
specific values of $Q_R$ for which $G^{\rm tot}(Q_R)=0$. Thus, one can also exploit
the AC ring in schemes where the spin-polarization of fully polarized injected current is
switched to the opposite direction via external electric field applied  through a gate
electrode covering the ring.~\cite{switch} For quantum-interference effects, it is important to
note that the purity of transported spins $|\vect{P}|=1$ remains one. Therefore, preservation of
full quantum coherence ensures complete visibility of the destructive
spin-interference effects that give rise to $G^{\rm tot}(Q_R)=0$.

Furthermore, our framework makes it possible to study single-channel transport through 2D ring, as
shown in Figure~\ref{fig:2d_single_total} which plots the $Q_{\rm R}$-dependence of
$G^{\rm tot}$ for the finite-width rings $(M,N)=(3,200)$ and at different Fermi energies selected
by the vertical arrows (a)--(c) in Fig.~\ref{fig:2d_multi_energy}. At these values $E_F$,
only one conducting channel
is available for quantum transport. We emphasize that the single-channel transport in finite width
conductors $M \geq 2$ is not equivalent to the transport in strictly 1D rings~\cite{ring,diego,molnar}
since the presence of unoccupied modes (evanescent modes) can influence the transport flowing through
the open channel in a way which depends on the confinement potential and
the geometry of the conductor.~\cite{decoherence,hausler} While Fig.~\ref{fig:2d_single_total} shows
that the calculated conductance still exhibits zeros at approximately the same values of $Q_{\rm R}$
as in the strictly 1D case,~\cite{diego} we emphasize that such zeros can be washed out
for transport occurring at specific Fermi energies (see also Fig.~\ref{fig:visibility}).
Also, the conductance oscillation patterns are rather irregular compared with the strictly 1D case,
especially in the
large $Q_{\rm R}$ regime.

\section{Visibility of spin-interference effects in multichannel quantum transport
through AC rings} \label{sec:multich}

\subsection{Injecting current through spin-polarized conducting channels} \label{sec:a}

In Fig.~\ref{fig:2d_multi_total} we show the conductance of the finite width ring $M=3$
for various Fermi energies which are indicated by vertical arrows (d)--(g) in
Fig.~\ref{fig:2d_multi_energy}. More than one transverse propagating mode in the
leads exist at these Fermi energies. Therefore, one can view the injected current
as being comprised  of electrons prepared in all of different quantum state
$|p \rangle \otimes |\sigma\rangle$ ($p \le M$). For non-magnetic leads, both $|p \rangle
\otimes |\!\! \uparrow$ and $|p \rangle \otimes |\!\! \downarrow$ electrons are injected
into the ring. The number of the conducting channels $M_{\rm open} \le M$ is denoted in
the Figure. Although conductance continues to display oscillating behavior as a function of
$Q_R$ even in the multichannel transport, its pattern is rather different from the single
channel case due to lack of $G_R(Q_R) \simeq 0$ points. Furthermore, the conductance
oscillation pattern is significantly more sensitive to the Fermi energy than in the single channel case.

The spin-resolved conductances in Fig.~\ref{fig:2d_multi_spin_1} for a given Fermi energy $E_F=-0.1t_0$
(at which $M_{\rm open}=3$) provide a more detailed information about such "incomplete" conductance
modulation. Here we consider the conductances corresponding to the injection of spin-$\up$ polarized
current. Similarly to the case of strictly 1D ring, the spin-conserved conductance $G^{\up\up}$
decays while the spin-flipped conductance $G^{\dn\up}$ increases as increasing $Q_{\rm R}$. Their
sum $G^{\up\up}+G^{\dn\up}$ is just half of the total conductance $G^{\rm tot}$ plotted in
Fig.~\ref{fig:2d_multi_total}. Despite the fact that multichannel AC rings are not able to
modulate (i.e., $G^{\rm tot}(Q_R) \neq 0$ at any $Q_R$) unpolarized current to the extent found
in 1D rings (where $G^{\rm tot}(Q_R) = 0$ at specific values of $Q_R$), the properties of current
spin-polarization vector in Fig.~\ref{fig:2d_multi_spin_1} suggest that they can serve as, even
better than 1D rings,~\cite{switch} spin-switch devices. That is, at large $Q_R$ such device
flips the spin-$\up$ of an incoming electron in the left lead into spin-$\dn$ of the outgoing
electron in the right lead.

The most prominent distinction between the single and the multichannel cases is that
the modulus of the spin polarization $|\vect{P}|$ can drop below one---the spin state
injected into the multichannel ring loses its purity for $Q_R \neq 0$. This is very
contrastive to the single channel transport case where the $|\vect{P}|=1$ is exactly
satisfied at any $Q_R$. Such a reduction of $|\vect{P}|$ is attributed to the fact
that finite Rashba SO coupling can induce  entanglement between the spin state of transported
electron and its orbital state, leaving a spin in a mixed quantum state
$\hat{\rho}_s^2 \neq \hat{\rho}_s$.

The spin-resolved conductances and spin polarization vector behave rather differently
depending on the polarization of the injected spin, as demonstrated by comparing
Fig.~\ref{fig:2d_multi_spin_1} and Fig.~\ref{fig:2d_multi_spin_2}, where we change
the direction of polarization of injected spin-$\up$ current to lie along the $y$-axis.
In this case, both the spin conserved conductance $G^{\up\up}$ and the spin flipped conductance
$G^{\dn\up}$ oscillate and contribute to the total conductance on any interval of $Q_{\rm R}$.
This is because the polarization of injected current in this case is orthogonal to the direction of
$\vect{B}_{\rm Rashba}(\vect{k})$ field at the interface between the left lead and the ring
(see Fig.~\ref{fig:ring}). Thus, the injected spin will be transported through the ring as a
superposition of $\up$ and $\dn$ spin states along the radial direction which causes the
oscillations of $G^{\dn\up}$ and $G^{\up\up}$. When the $y$-polarized current is injected,
$P_x$ and $P_z$ characterizing the spin current are close to zero, while $P_y$ exhibits
quasi-periodic oscillation (note that, according to  Eq.~(\ref{eq:Pvector}) oscillations
in $P_y$ are in one-to-one correspondence with oscillations of $G^{\dn\up}$ and $G^{\up\up}$).
Since $P_x, P_z \simeq 0$, the purity of the transported spin state is approximately
given by $|\vect{P}| \approx |P_y|$.

\subsection{Injecting current through eigenchannels} \label{sec:b}
\begin{figure}[ht]
\begin{center}
\vspace*{-0.5cm}
{\includegraphics[scale=.3]{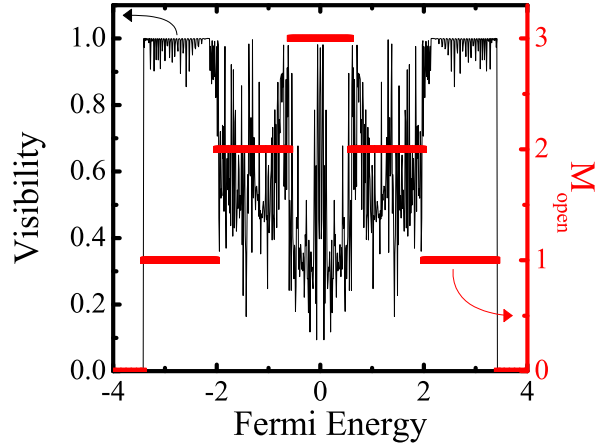}}
\vspace*{-0.5cm}
\end{center}
\caption{The visibility of spin-interference effects, defined by
Eq.~(\ref{eq:visibility}), for a finite width ring conductor $(M,N)=(3,200)$.
The number of the open orbital channels $M_{\rm open}$ is indicated on the
right $y$-axis of the plot.}
\label{fig:visibility}
\end{figure}

In strictly 1D rings (or transport through a single open channel of a 2D ring) the
conductance goes to zero $G^{\rm tot}=0$ (or $G^{\rm tot} \simeq 0$, see
Fig.~\ref{fig:visibility} in the regime $M_{\rm open}=1$) at specific values of the
Rashba SO coupling due to destructive interference effects in coherent superpositions
$a  |\!\! \uparrow \rangle + b |\!\! \downarrow \rangle$, as discussed in Sec.~\ref{sec:onech}.
When current is injected also through higher transverse modes, the analysis of the ring transport
properties becomes much more involved. Nevertheless, one can envisage unfolding of three plausible
scenarios:

\begin{enumerate}

\item The difference in AC phases acquired by spin states is the upper and lower
branches of the ring is independent of the wave vector of transverse modes, so that electron
remains in a separable quantum state $\left (\sum_{p=1}^M c_p |p \rangle \right)
\otimes (a  |\!\! \uparrow \rangle + b |\!\! \downarrow \rangle)$
(i.e., the transmission matrix can be decoupled into a tensor product of spin-dependent part and
a spatial scattering part~\cite{entin}); in this case the conditions for destructive interference
remain the same as in single-channel quantum transport,~\cite{ring}

\item The transmitted spins through different channels, each of which has its own orbital phase,
pick up different AC phases so that conditions for a destructive interference $G^{\rm tot}=0$
cannot be satisfied simultaneously in all channels,

\item The transmitted spins lose their coherence due to coupling to environment, i.e.,
the off-diagonal elements of the spin density matrix are decaying in the course of
transport;~\cite{decoherence} if they do not diminish all the way to zero, the
transported spin will end up in a {\em partially coherent} quantum state.~\cite{gefen,partial_fano}

In both the first and the second scenario, the spin is described by a pure quantum state in
the course of transport. Although the second and the third scenario lead
to the same observable consequences---the ring conductance never reaches zero
since visibility~\cite{gefen} of the spin-interference effects is
reduced below one---they are fundamentally different. In the second scenario, spin states remain fully coherent, while in the third one transmitted spins are partially coherent due to coupling to the environment. Even when all other (usually many-body~\cite{partial_fano}) decoherence mechanisms are suppressed,
single spin of an electron can still be entangled to the environment composed
of its orbital channels. This mechanism becomes operable in the clean ring when spin-independent
charge scattering off boundaries and interfaces occurs in the presence of SO coupling.~\cite{decoherence}
The partially coherent states, as an outcome of entanglement of spin of transmitted electron with the spin in a quantum dot, were also found  in recent experiments on Aharonov-Bohm ring interferometers where quantum dot is embedded in  one ring arms.~\cite{partial_fano}

Here we explore  possibility of the same  partially coherent outgoing spin state to appear in
the AC ring, where physical mechanism of entanglement is different and single-particle in nature.~\cite{decoherence} This makes investigation of ring transport properties  in terms of
\end{enumerate}

the AC phases acquired by circulating spins rather difficult since one has to extract geometric phase of an open spin quantum system described by the density matrix rather than by a pure state.~\cite{vedral} It is insightful to define the {\em visibility} of quantum interference
effects in the AC ring
\begin{equation}\label{eq:visibility}
{\mathcal V}(E_F) = \frac{G_{\rm max}^{\rm tot}(E_F) - G_{\rm min}^{\rm tot}(E_F)}{G_{\rm max}^{\rm tot}(E_F)}.
\end{equation}
Here $G_{\rm max}^{\rm tot}$ and $G_{\rm min}^{\rm tot}$ are the maximum and the minimum
values of the total conductance found in the first period of the conductance oscillations
vs. $Q_R$, for a given Fermi energy and corresponding number of open channels
$M_{\rm open} < M$. The visibility ${\mathcal V}(E_F)$ in 2D rings allowing for a maximum
of three open channels is plotted in Fig.~\ref{fig:visibility}, which shows that
${\mathcal V}(E_F) \approx 1$ in the single-channel transport regime, i.e., the unpolarized charge transport can be fully modulated by changing the strength of the Rashba SO interaction
(except at particular Fermi energies). However, as soon as the second channel starts
contributing to the transport, ${\mathcal V}(E_F)$ decreases below one.
\begin{figure}[ht]
\begin{center}
\vspace*{-0.5cm}
{\includegraphics[scale=.32]{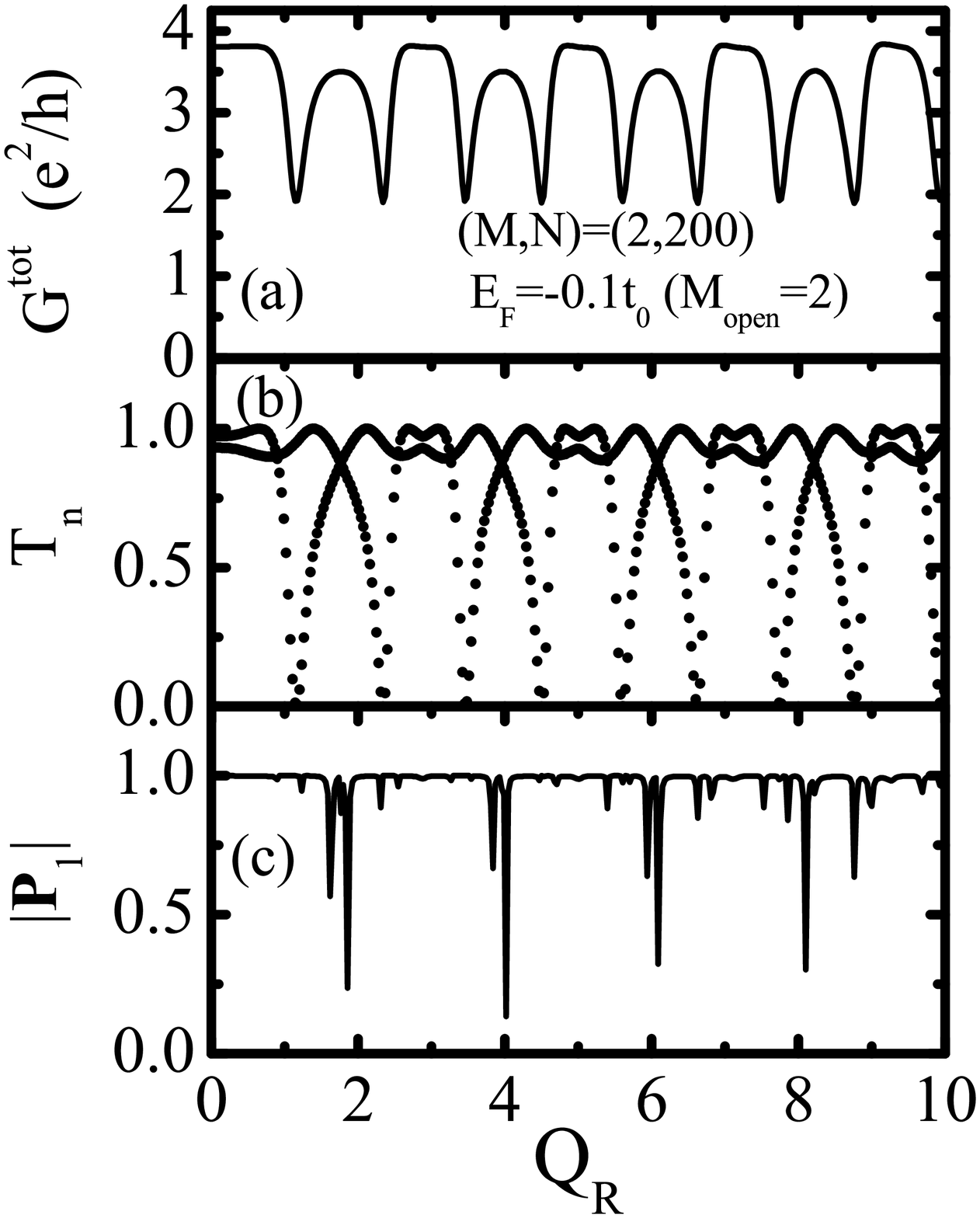}}
\vspace*{-0.5cm}
\end{center}
\begin{center}
\vspace*{-0.5cm}
{\includegraphics[scale=.32]{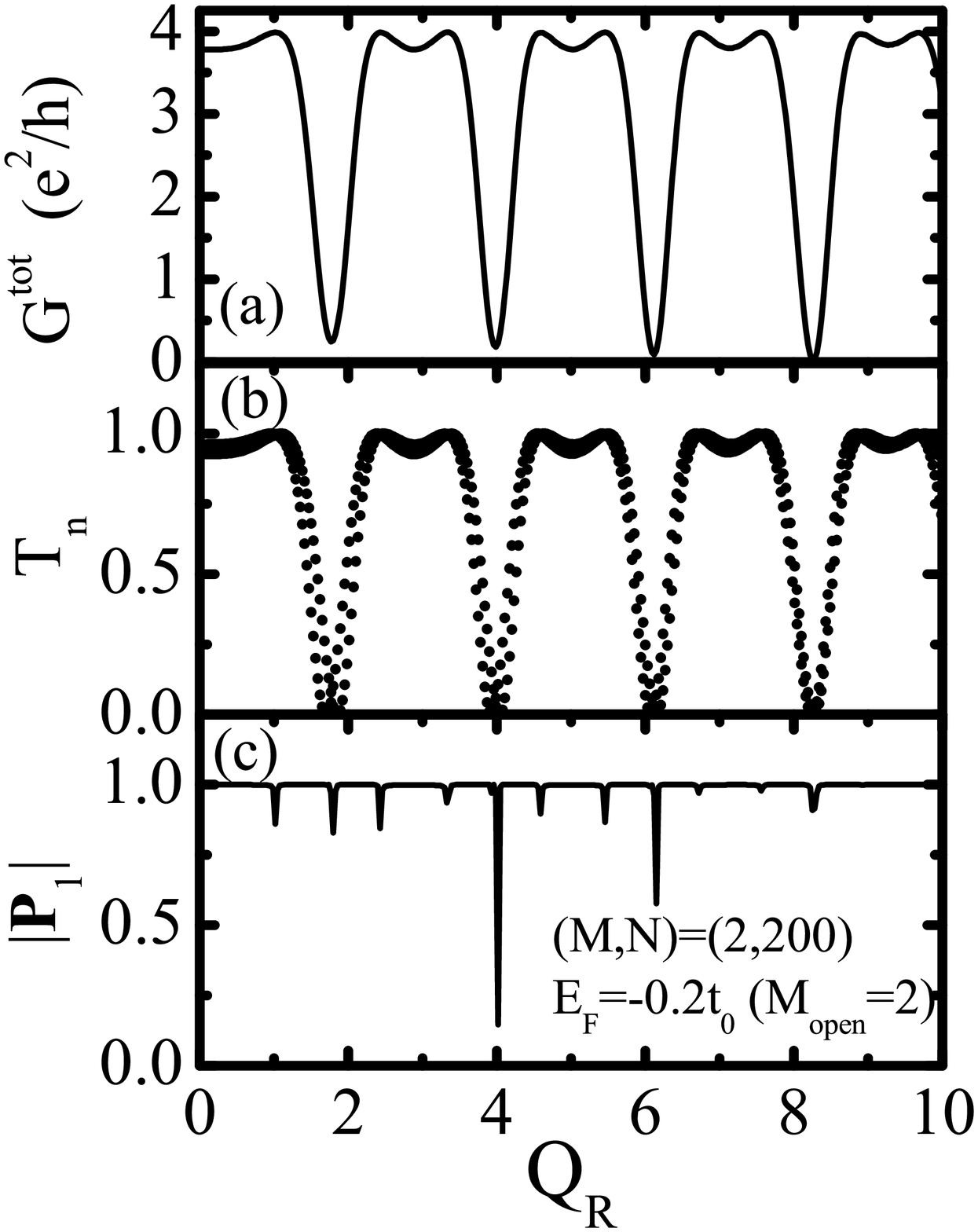}}
\vspace*{-0.5cm}
\end{center}
\caption{The total conductance (a) $G^{\rm tot}(Q_R,E_F)$, (b)
transmission eigenvalues $T_n(Q_R,E_F)$, and (c) modulus of spin-polarization
 vector $|\vect{P}_1|(Q_R,E_F)$ corresponding to the first eigenchannel
of a {\em two-channel} ring conductor with $(M,N)=(2,200)$ and at the Fermi energies
$E_F=-0.1t_0$ (upper panel) or $E_F=-0.2t_0$ (lower panel). Note that due to
Kramers degeneracy (in the presence of SO interaction, but absence of magnetic
fields, rings Hamiltonian Eq.~(\ref{eq:ring_hamil}) is time-reversal invariant) there are
only two different $T_n$ at each $Q_R$.}
\label{fig:eigen_2ch}
\end{figure}
\begin{figure}[ht]
\begin{center}
\vspace*{-0.5cm}
{\includegraphics[scale=.35]{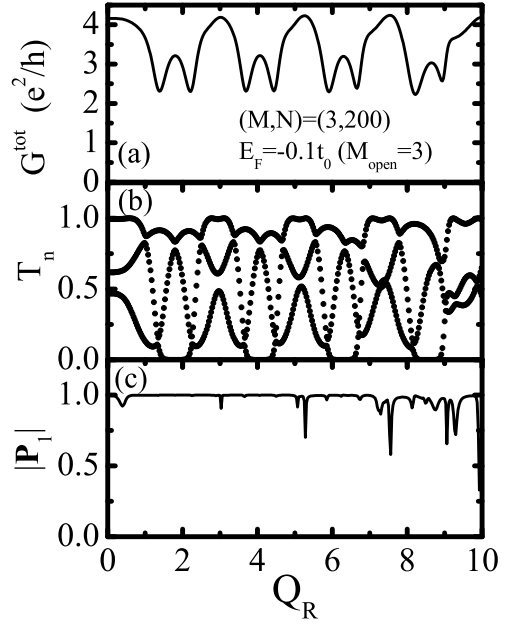}}
\vspace*{-0.5cm}
\end{center}
\caption{The total conductance (a) $G^{\rm tot}(Q_R,E_F)$, (b)
transmission eigenvalues $T_n(Q_R,E_F)$, and (c) modulus of spin-polarization
 vector $|\vect{P}_1|(Q_R,E_F)$ corresponding to the first eigenchannel
of a {\em three-channel} ring conductor $(M,N)=(3,200)$ and at the Fermi energies
$E_F=-0.1t_0$ (upper panel) or $E_F=-0.2t_0$ (lower panel). Due to the Kramers
degeneracy there are only three distinctive points plotted at each $Q_R$.}
\label{fig:eigen_3ch}
\end{figure}

To investigate which of the above scenarios is realized in the transport through
multichannel rings, it is advantageous to invoke as much as possible transparent
picture of spin-interference effects in 1D rings.~\cite{ring,diego,molnar} Thus,
to be able to consider the transport through multichannel ring as if taking place
through a system of independent single-channel rings, we switch to the representation
of eigenchannels. In general, the basis of eigenchannels, in which ${\bf t} {\bf t}^\dag$
is a diagonal matrix, offers a simple intuitive picture of the transport in  a mesoscopic
conductor that can be viewed as a parallel circuit of independent channels characterized
by channel-dependent transmission probabilities $T_n$. The computation of $T_n$ as
eigenvalues of ${\bf t} {\bf t}^\dag$ in the case of conventional unpolarized charge transport allows one to obtain a plethora of transport quantities beyond just the conductance through simple expressions.~\cite{carlo_rmt} However, the rotation to the diagonal ${\bf t} {\bf t}^\dag$ matrix is inapplicable~\cite{decoherence} in spintronics where usually the spin density matrix of injected electrons is non-trivial in the incoming channels of the leads. Nevertheless, when both spin-$\uparrow$ and spin-$\downarrow$ are injected into the ring in equal proportion, the basis of eigenchannels allows us to "deconstruct" 2D ring into $M_{\rm open}$ single-channel rings. Moreover, in systems with SO interaction, which are
time-reversal invariant in the absence of external magnetic field, due to the
Kramers degeneracy all transmission eigenvalues are double degenerate.~\cite{carlo_rmt}
In the case of single-channel rings of Sec.~\ref{sec:onech} this correspond to both spin-$\uparrow$ and spin-$\downarrow$ electrons giving identical contribution to $G^{\rm tot}$. The observable transport properties of a multichannel ring cannot differentiate between injection of unpolarized current through spin-polarized channels defined by the leads as a boundary condition (which were utilized in Sec.~\ref{sec:a}) or through the eigenchannels.

The spin properties of eigenstates of ${\bf t} {\bf t}^\dag$ can be described by assigning
the density matrix to each eigenchannel $|n \rangle = \sum_{p,\sigma} \epsilon_{p,\sigma}^n
|p \rangle \otimes |\sigma \rangle$
\begin{equation}
\hat{\rho}^n=\sum_{p p^\prime \sigma \sigma^\prime} \epsilon_{p,\sigma}^n
\epsilon_{p^\prime,\sigma^\prime}^{n*}   |p \rangle \langle p^\prime|
\otimes |\sigma \rangle \langle
\sigma^\prime|,
\end{equation}
and then taking the partial trace over the orbital degrees of freedom to get
the reduced density matrix for the spin subsystem of an eigenchannel
\begin{equation}
\hat{\rho}^n_s=\sum_{p \sigma \sigma^\prime} \epsilon_{p,\sigma}^n
\epsilon_{p,\sigma^\prime}^{n*}  |\sigma \rangle
\langle \sigma^\prime|.
\end{equation}
This allows us to extract the purity $|{\bf P}_n|$ of the spin subsystem of an
eigenchannel from ${\bf P}_n={\rm Tr}\, [\hat{\rho}_s^n \hat{\bm \sigma}]$.

Figure~\ref{fig:eigen_2ch} shows the total conductance, the eigenchannel transmissions,
and the spin purity $|\vect{P}|$ of an eigenchannel of the two-channel $(M,N)=(2,200)$
ring conductor at two different values of the Fermi energy ensuring that both conducting channels 
in the leads are open for transport. Here we plot only the spin purity $|\vect{P}_1|$ 
corresponding to the first eigenchannel (spin purity of other eigenchannels display similar behavior). 
Since the eigenchannel transmissivities are twofold degenerate,
one can observer only two different values of $T_n$ at each $Q_{\rm R}$ in Fig.~\ref{fig:eigen_2ch}. 
When $E_F=-0.1t_0$, the total conductance shows incomplete modulation ${\mathcal V} \simeq 0.5$ 
since the conductance never reaches zero value. 
It is possible to recognize in Fig.~\ref{fig:eigen_2ch} that the eigenchannel
transmissions form two distinctive curves as a function of $Q_R$. Furthermore, each of
them exhibits full-modulation characterized by $T_n=0$ at particular values of
$Q_R$. However, these two oscillating patterns are shifted with respect to each other,
thereby preventing total conductance $G^{\rm tot}(Q_R) = 2e^2/h \, [T_1(Q_R) + T_2(Q_R)] > 0$
from reaching zero value at any strength of the SO interaction. This can be explained as
a realization of the second scenario introduced above. On the other hand, at $E_F=-0.25t_0$
the total conductance shows almost complete modulation ${\mathcal V} \simeq 1$ because: (i)
the individual eigenchannel transmissions are akin to the ones found in single-channel rings;
and (ii) they almost completely overlap with each other. This case, albeit found rarely
in multichannel AC rings, represents quite a good example of the first scenario mechanism. Interestingly enough, $|{\bf P}_1|$ drops below one (meaning that an electron is ``injected''
into the conductor in an entangled state of spin and orbital channels) at those values
of $Q_R$ where one would expect destructive interference effects of pure spin states in single-channel rings or system of such independent rings.

Figure~\ref{fig:eigen_3ch} plots the same eigenchannel physical quantities for the
transport through a three-channel ring  $(M,N)=(3,200)$, where $E_F=-0.1t_0$ is set
to allow all three channels to be opened. The total conductance in this case also
displays incomplete modulation ${\mathcal V} \simeq 0.3$. However, in this case only
one $T_n(Q_R)$ curve can isolated that oscillates between 0 and 1, while the other
two never reach zero values. Moreover, these three patterns of $T_n(Q_R)$ are shifted
with respect to each other. This observation within the picture of three independent
single channels explains why the oscillations of the total conductance end up having
a rather small amplitude $G^{\rm tot}_{\rm max}$ - $G^{\rm tot}_{\rm max}$.

\section{Conclusion} \label{sec:conclusion}

In conclusion, we have investigated how the conductance of unpolarized electron
transport through two-dimensional rings changes as we increase the strength of
the Rashba SO coupling. Moreover, we connect the properties of the charge transport
to the orientation of spin of injected electrons as well as its coherence properties.
In order to take into account the effect the finite width of the ring and the leads systematically, 
we model the ring using concentric tight-binding lattice Hamiltonian
as the starting point for the  calculation of charge and spin transport properties
based using the Landauer transmission matrix for a two-probe device.

Our analysis suggests that conductance oscillations, induced by changing the
Rashba SO coupling, will persist to some extent even in the multichannel transport
through mesoscopic ring-shaped conductors. However, the oscillation patterns are
rather different from the single-channel case, or from the anticipated oscillations
in multichannel rings where spin-interference effects would be equivalent in all
channels and simply add up. The conductance of single-channel transport through the
ring displays full modulation, where $G^{\rm tot}(Q_R)=0$ appears quasi-periodically
as a function of the Rashba SO coupling. This effect is explained in simple terms~\cite{ring,diego,molnar} 
as a result of  destructive spin-interference effects
between opposite spin states circulating in the clockwise and counterclockwise direction around the ring, 
where they acquire Aharonov-Casher phase in the presence of
the Rashba electric field. On the other hand, quantum transport in most of the
cases occurring in multichannel rings does not lead to zero conductance at any
Rashba coupling. Using the picture of quantum transport through independent eigenchannels,
we identify different scenarios washing out $G^{\rm tot}(Q_R)=0$ feature of
the single-channel transport, which involve either pure or mixed transported
spin states. As the number of conducting channels increases, it becomes more likely
to generate  partially coherent transported spin (which is described by the density
matrix rather than the pure state vector) due to entanglement of electron spin to its
orbital degrees of freedom.~\cite{decoherence}

\begin{acknowledgments}

We thank F. E. Meijer, J. Nitta, and M. Shayegan, and D. Pappas for sharing with
us important experimental insights.

\end{acknowledgments}



\end{document}